\documentstyle[12pt]{article}
\begin{document}
\centerline{\bf Symplectic dynamics of the nuclear mean-field\footnote{
in  Proc. Int. School {\it Collective Motion and Nuclear Dynamics}, Predeal, Romania, 1995, 
Edited by A. A. Raduta, D. S. Delion, I. I. Ursu, World Scientific, Singapore, 1996, p. 251-259.}} 
\vspace{1.5cm}
\centerline{ M. Grigorescu}
\vspace{1.5cm}
\noindent
$\underline{~~~~~~~~~~~~~~~~~~~~~~~~~~~~~~~~~~~~~~~~~~~~~~~~~~~~~~~~
~~~~~~~~~~~~~~~~~~~~~~~~~~~~~~~~~~~~~~~~~~~}$ \\[.3cm]
Collective and microscopic pictures of nuclear dynamics are related in the framework of time-dependent variational principle on symplectic trial manifolds. For symmetry breaking systems such manifolds are constructed by cranking, and applied to study the nuclear isovector collective
excitations. \\
$\underline{~~~~~~~~~~~~~~~~~~~~~~~~~~~~~~~~~~~~~~~~~~~~~~~~~~~~~~~~
~~~~~~~~~~~~~~~~~~~~~~~~~~~~~~~~~~~~~~~~~~~}$ \\[1cm]
{\bf PACS:} 21.60.-n, 21.60.Fw, 03.75.Kk   \\[3cm]

\newpage
\section{Introduction} 
The interplay between quantum and classical aspects in the nuclear collective dynamics is a long standing puzzle, similar to the one represented by the quantum behaviour of a macroscopic variable \cite{leg}, and is not yet completely understood. \\ \indent
The phenomenological collective models are based essentially on the quantization of some simple 
classical systems (rigid body, liquid drop), with parameters obtained by fitting data \cite{bm}. Therefore, in this approach the nucleus is supposed to be almost frozen, because the
model Hamiltonian depends on a small number of variables. Also,
the Hilbert space constructed by quantization is rather
artificial, because it does not account for all the observables
of the many-nucleon system. \\ \indent
In the microscopical approach, the nuclear dynamics is
represented in the full Hilbert space by the time-dependent Hartree-Fock
(HF), or Hartree-Fock-Bogolyubov (HFB) equations \cite{rs}. These
equations have static solutions corresponding to the ground
state, while the excited states are obtained by the random phase
approximation (RPA), or more generally, using
requantization methods \cite{gipq,grig1}. A clear sign of
collective behaviour, in the sense of the phenomenological
models, appears when the ground state is non-invariant to a
continuous symmetry group of the Hamiltonian (spontaneous symmetry
breaking). This situation may appear for geometrical symmetries,
like rotation, or dynamical, like particle-number,
and the corresponding nuclei are known as deformed, respectively
superfluid. For these nuclei, the inertial parameters of the collective
motion may be calculated microscopically using the average
energy of the ground state, shifted to a "moving frame" by the symmetry
generators (the "cranking" method). However, the cranking 
is not related strictly to the symmetry breaking and is applied,
for instance, to obtain the effective mass at
fission \cite{brack}. \\ \indent
The choice of the cranking operators is not arbitrary, because they should
be in one-to-one correspondence with the canonical collective variables,
up to a unitary change in the Hilbert space representation.  
This problem of correspondence is not simple, but a natural
solution can be given if the cranking is applied at a more
fundamental level, to generate both the collective phase
space and the phenomenological Hamiltonian. 
The key object necessary for this purpose is represented by the
shifted ground states, which play only a secondary role in the standard
cranking calculation. These states are parameterized by the "shift"
variables, and may be joined in trial manifolds ({\it S}), endowed with a classical phase space structure (symplectic form) induced from the many-body Hilbert space \cite{grig2}. 
By construction, {\it S} should be considered both as collective phase
space and as trial manifold  for a time-dependent variational
treatment of the microscopic Hamiltonian. Moreover, the Hamilton equations of motion
for the collective variables are related kinematically to the
evolution of the trial functions produced by the variational calculation.  
Therefore, the "artificial" quantization of the collective
motion becomes useless, and can be replaced by requantization, to
obtain directly the microscopic states which correspond to a particular
collective motion.  \\ \indent
It is interesting to remark that the standard RPA is included in
this formalism as a limit case, reached when {\it S} is constructed
without any correspondence to an intuitive model, but taking as generators
all possible particle-hole (ph) or particle-particle (pp) operators
in some finite basis. This nice example will be discussed in the
next section. Then, Section 3 presents the  
symplectic trial manifolds for deformed and for superfluid systems,
and two applications to the treatment of the isovector "scissors-like" 
collective excitations. 
\section{Symplectic dynamics on trial manifolds and the RPA} 
\subsection{Symplectic dynamics and requantization }
Let us assume that ${\cal H}$ is the many-body Hilbert space, $H$
is the microscopic  Hamiltonian,  
${\it S}= \{ \vert \psi \rangle (X) \}$ is a $2N$-dimensional trial manifold
of normalized functions, parameterized by the variables
$X = \{ x^i \}$, $i=1, 2N$, and that the matrix $ \omega^{\it S} =
[ \omega_{ij}^{\it S} ( \psi)]$,  
$$\omega_{ij}^{\it S} ( \psi) = 2 \hbar Im \langle  \partial_i \psi \vert \partial_j \psi \rangle ~~, $$ 
is non-singular.
Thus, $ \omega^{\it S}$ defines a symplectic form on {\it S}, 
and the functional 
$${\cal J}[X] = \int_a^b \langle  \psi \vert i \hbar \partial_t - H \vert \psi \rangle dt $$ 
is stationary for the solution $X_t$ of the Hamilton equations
\begin{equation}
 \sum_{j=1}^{2N} \dot{x}^j \omega_{jk}^{\it S}(\psi) = 
\frac{\partial \langle  \psi \vert H \vert \psi \rangle}{ \partial x^{k}}~~.
\end{equation}
Therefore, the solution of a time-dependent variational calculation
within {\it S} has the form $\vert \psi \rangle (X_t)$, where $X_t$ is a trajectory
given by Eq. (1). \\ \indent
It is interesting to note that if the couple ${\cal H}$, $H$ corresponds to the quantum harmonic oscillator, and  {\it S}  is the manifold of the Glauber coherent
states, then (1) represents the Hamilton equations for the classical oscillator. \\ \indent
The procedure of extracting information about the spectrum of
$H$ from the orbits $\vert \psi\rangle(X_t)$ is called
"requantization", and if the system of Eq. (1) is integrable, then the
method GIPQ \cite{gipq} (gauge-invariant periodic quantization) can be 
applied. According to this method, the periodic orbits $\gamma = \{ X_t^\gamma \}$,
$$  X_t^\gamma = X_{t+T_\gamma}^\gamma~~, $$ 
should be quantized by a Bohr-Wilson-Sommerfeld (BWS) condition
\begin{equation}
\int_0^{T_\gamma} dt \langle  \psi \vert i \partial_t \vert \psi\rangle = 2 \pi n~~,
~~n=0,1,2,...~ .
\end{equation}
This gives  the energy spectrum, while the eigenstates corresponding
to a quantized orbit $\gamma^n$ should be approximated by the time-average \cite{grig1}
\begin{equation}
\vert \Omega_\gamma \rangle= \frac{1}{T_{\gamma^n}} \int_0^{T_{\gamma^n}} 
dt e^{i \Theta_t} \vert \psi\rangle(X_t^{\gamma^n})
\end{equation}
with $\Theta_t  = \int_0^t dt' \langle  \psi \vert i \partial_{t'} \vert
\psi\rangle $ (the "Berry's phase").  \\ \indent
An extended version of GIPQ includes also the quantization of
the invariant tori \cite{gipq2}. As it will be shown further, the physical 
applications strongly support the requantization by Eq. (3), though its geometrical
meaning is not yet completely clear. The study of such integral representations
("quantization by membranes") represents an active field of research   
in mathematical physics \cite{kara2}.         
\subsection{The random phase approximation } 
We suppose now that the output of a static mean-field calculation
(HF or HFB) provides the ground state $\vert g\rangle$ and a set
of $2N$ operators $E_{\pm \alpha}$, $\alpha =1,N$, $E_{-\alpha}= E_\alpha^\dagger$,
so that $E_{-\alpha} \vert g\rangle = 0$. If the single-particle 
basis contains n states, then in the HF case
$\vert g\rangle \equiv c^\dagger_1 c^\dagger_2 c^\dagger_3 ... c^\dagger_A \vert 0\rangle $
is constructed by acting with $A$ fermion creation operators
$c^\dagger_h$, $h=1,A$, on the particle vacuum $\vert 0\rangle$, 
and a possible choice is $E_\alpha=c^\dagger_pc_h$, with $1 \le
h \le A$, $A <  p \le n$, and $N=A(n-A)$. If there are no "hole" states, and $\vert g \rangle \equiv \vert 0\rangle $ is the particle (or quasiparticle in HFB) vacuum, then we may 
have $E_\alpha = c^\dagger_{p_i} c^\dagger_{p_j}$, with $p_i < p_j = 1,n$, and 
$N=n(n-1)/2$. \\ \indent
The operators $E_{\pm \alpha}$ can be used to generate a trial manifold 
${\it S}^{RPA}$  represented by the set of functions
\begin{equation}
\vert \psi \rangle(Z) \equiv U(Z) \vert g \rangle~~,
\end{equation}
where
\begin{equation}
U(Z) = e^{\sum_{\alpha} (z_\alpha E_\alpha - z_\alpha^* E_{- \alpha})}
\end{equation}
is an unitary operator and $z_\alpha$ are $N$ complex variables. Therefore ${\it S}^{RPA}$ is parameterized by $2N$ real variables, denoted $x^i$, $i=1,2N$, so that for $i \le N$, $x^i$ are
$Re(z_\alpha)$, and for $N< i \le 2N$, $x^i$ are 
$Im(z_\alpha)$. \\ \indent
The condition as $\vert g\rangle$, (or $x^i = 0$), to be a fixed point for
Eq. (1) gives the static "mean-field" equations 
$$\langle g
\vert[H,E_{- \alpha}] \vert g\rangle =0~~,$$
automatically fulfilled in the HF or HFB case. \\ \indent
Let us consider now a small amplitude vibrational periodic orbit $\gamma$ around $\vert g\rangle$
with the period $T$, so that all $x^k$ perform harmonic oscillations. This means 
$$z_\alpha = X_\alpha e^{-i \Omega t} + Y_\alpha e^{i
\Omega t}~~,$$ 
with $\Omega= 2 \pi /T$, and $T$, $X_\alpha$, $Y_\alpha$  unknowns which
should be determined from the equations of motion. For the orbit
$\gamma$,
\begin{equation}
U ( Z^\gamma_t) = \exp( e^{-i \Omega t} B^\dagger - e^{i \Omega t} B)
\end{equation}
with 
$$
B^\dagger = \sum_{\alpha } (X_\alpha E_\alpha - Y_\alpha^* E_{-\alpha})~~,
$$
and in the linear approximation Eq. (1) reduces to the RPA-type equation      
\begin{equation}
\langle g \vert [[H,B^\dagger] - \hbar \Omega B^\dagger, E_{\pm \alpha}] \vert g\rangle =0~~.
\end{equation}
The standard particle-hole or particle-particle RPA are recovered when $\vert g\rangle$ is the HF or HFB ground state and $E_\alpha=c^\dagger_pc_h$, respectively $E_\alpha = c^\dagger_{p_i} 
c^\dagger_{p_j}$. From Eq. (7) one obtains the "normal mode frequency" $\Omega$ and a
one-parameter family of periodic orbits having this frequency. As a parameter we may consider for instance the energy $${\cal  E}=\langle  g  \vert U(Z)^{-1} H U(Z) \vert g\rangle ~~.$$ 
However, if ${\cal E}$ is too large, the amplitudes $X,Y$
increase and Eq. (7)  fails in approximating Eq. (1). \\ \indent
Until now the whole discussion was about the classical dynamics on
${\it S}^{RPA}$. To establish the connection with the quantum many-body
system is necessary to requantize the RPA periodic orbits
according to Eq. (2). 
To perform this integral we will assume that exists a Hermitian operator
$W$, so that $[W,B^\dagger] = B^\dagger$, although the explicit form of W
is not necessary. If $W$ exists, then
 $$U^{-1} i \partial_t U = \Omega (U^{-1} W U -W) =
\Omega ( e^{-i \Omega t} B^\dagger + e^{i \Omega t} B + [B,B^\dagger] +
...)~~.$$ 
By integrating this sum over a period, the terms
linear in $B, B^\dagger$ vanish, and the first non-vanishing term is 
$2 \pi [B,B^\dagger]$. Thus, the  BWS quantization gives 
$$\langle g \vert [B,B^\dagger] \vert g\rangle = n~~.$$ 
For $n=1$ (the first excited state), this coincides with the RPA 
"normalization" condition, an interesting result proved before using path 
integrals \cite{negele}.
\\ \indent
 After quantization,  the state associated by Eq. (3) to $\gamma$
can be easily obtained expanding $U$ in powers of $B^\dagger,
B$, and retaining only the linear terms. 
The result has the familiar form $$\vert \Omega\rangle = B^\dagger \vert g\rangle~~, $$ 
but by contrast to the RPA assumption, the excitation 
operator $B^\dagger$ acts on  the uncorrelated ground state. However, 
on particular examples it can be shown that if the whole expansion
of $U$  is considered, then $ \vert \Omega\rangle=B^\dagger P_{RPA} \vert g\rangle$, with 
$P_{RPA}$ a Hermitian operator which gives an approximate projection of $\vert g\rangle$ on
the vacuum $\vert RPA\rangle$ of $B$, defined by $B \vert RPA\rangle=0$. Moreover,
if $\vert g\rangle$ is symmetry breaking and $\gamma$ is a related rotational orbit, 
then  Eq. (3) gives $P_{s} \vert g\rangle$, with $P_s$
a symmetry restoring projection operator \cite{grig1}.     
\section{Isovector excitations in symmetry breaking nuclei} 
\subsection{The two rotor model } 
The prediction of the isovector angular rotational 
oscillations \cite{iudice} (scissors vibrations) in deformed nuclei
has been particularly stimulating for the experimental research
on the nuclear magnetism, leading to the discovery of low-lying 
M1 states. These states have been observed in high resolution $(e,e')$ scattering experiments on rare earths \cite{gd}, $fp$-shell  nuclei \cite{ti}, and in actinides \cite{act}. 
Their apparent weak excitation in intermediate energy proton scattering \cite{prot} has supported the orbital character predicted by the two rotor model (TRM), but the highly 
fragmented structure has generated a long standing debate about
their real origin. On one side were the phenomenological models 
supporting the TRM picture, like CSM \cite{rad}, 
or IBA-II \cite{iach}, while on the other were the microscopic
RPA or QRPA calculations, indicating that the observed M1 excitations
are produced by only few quasiparticle pairs. Not less important
for this debate was the difficulty to decide if the states obtained by microscopic calculations  correspond or not to angular vibrations. Therefore, the problem of finding the appropriate
microscopic correspondent for a specific collective motion appears to be important. \\ \indent 
This problem can be solved by an RPA calculation based on special
trial manifolds ${\it S}^{rot}$, generated by cranking, instead of
 ${\it S}^{RPA}$ defined in Section 2.2. 
Let us denote by ${\cal G}_{x}$ the group of rotations around 
the X axis, $L_x$ the orbital angular momentum operator, and by
$\vert g\rangle$ the axially-deformed ground state of the microscopic Hamiltonian $H$.
Then, the intrinsic ground state of a system rotating around the X-axis  with angular momentum ${\cal L}$ is given by the solution $\vert Z_\omega \rangle$ of the variational equation  
$$\delta \langle  Z \vert H - \omega L_{x} \vert Z \rangle=0~~. $$ 
The set of functions $\vert Z_\omega \rangle$ represents a curve in ${\cal H}$ containining $\vert g\rangle$ and parameterized by the Lagrange multiplier $\omega$, or, implicitly, by
the angular momentum 
$$ {\cal L} = \langle Z_\omega \vert L_{x} \vert Z_\omega \rangle~~.$$
The action of ${\cal G}_x$ moves this curve over a surface in ${\cal H}$ which contains the states
\begin{equation}
\vert \psi \rangle(\phi, \omega) = e^{-i \phi L_x / \hbar} \vert Z_\omega \rangle 
\end{equation}
and defines the trial manifold ${\it S}^{rot}$. \\ \indent
In arbitrary variables $\{ q, p \}$, the symplectic structure of ${\it S}^{rot}$ is given
by the 2-form $\omega^{rot}$,   
$$\omega^{rot}_{ q p} = 2 \hbar Im \langle  \partial_q \psi \vert \partial_p \psi \rangle~~. $$  An obvious choice of these variables is $ q = \phi$, and $ p$ 
a function of $\omega$. When this function is the angular momentum,  ($p={\cal L}$), then 
$$\omega^{rot}_{\phi {\cal L}} = 2 \hbar Im \langle  \partial_\phi \psi( \phi, \omega) 
\vert \partial_{\cal L} \psi( \phi, \omega)\rangle= \partial_{\cal L} \langle  Z_\omega  
\vert L_x \vert Z_\omega \rangle = 1~~,$$
proving that $\phi$ and ${\cal L}$ are canonical, and ${\it S}^{rot}$ is the phase space of the 
classical (plane) rotor. \\ \indent 
In the case of a deformed nucleus, ${\it S}^{rot}$ can be constructed
separately for protons and neutrons, and the trial wave function
corresponding to the total phase space ${\it S}_{pn}^{rot} = 
{\it S}_p^{rot} \times {\it S}_n^{rot}$ is
\begin{equation}
\vert \psi \rangle(\phi_p, \phi_n, \omega_p, \omega_n ) = 
e^{-i( \phi_p L_x^p + \phi_n L_x^n)/ \hbar} \vert Z^p \rangle_{\omega_p} 
 \vert Z^n \rangle_{\omega_n}~~.
\end{equation}
Let us consider now a schematic nuclear Hamiltonian
\begin{equation}
H =  \sum_{\mu, \nu} (h_0)_{\mu \nu} c^\dagger_\mu
c_\nu - \frac{ \chi_0}{2}( Q_{is} Q_{is}^\dagger + b
Q_{iv} Q_{iv}^\dagger)
\end{equation}
consisting of a spherical oscillator term ($h_0$ is the one-body spherical 
oscillator Hamiltonian with frequency $\omega_0= 41 A^{-1/3}$ MeV/$\hbar$),  
and the quadrupole-quadrupole (QQ) interaction, with both isoscalar and isovector components ($b \approx - 0.6$). Then, for ${\it S}_{pn}^{rot}$ and $H$, Eq. (1) takes the form of the Hamilton system of equations for two  rotors \cite{grig3}, having the cranking moments of inertia $I_p, I_n$,  and interacting by a restoring  elastic potential $C_\chi(\phi_p - \phi_n)^2/2$. 
Worth noting, $$C_\chi=3 \chi_0 (1-b) \langle Q^p_0\rangle_g \langle Q^n_0\rangle_g \approx 9 \delta^2 A {\rm ~~MeV} $$ 
appears related to the microscopic QQ interaction ($Q_{0}^{p,n}=\sqrt{{5}/{16 \pi}}
\sum_{p,n} (2 z^{2}-x^{2} -y^{2})_{p,n}$), by contrast to the TRM estimate\footnote{ F. Palumbo, "The scissors mode", in Proc. Int. School {\it Symmetries and semiclassical features of nuclear dynamics}, Poiana Brasov, Romania, 1986, Springer (1987), p. 230.}   
$$ C_{TRM} \approx 6 \delta^2 A^{4/3}~~{\rm MeV}~~, $$ related to the symmetry energy ($\delta$ is the deformation parameter)\footnote{$\delta = \beta \sqrt{45/ 16 \pi} $. If $\langle Q^p_0 \rangle_g = \langle Q^n_0\rangle_g $ then 
$\delta = 3 \chi_0 c_0 \langle Q_0^p \rangle_g / \hbar \omega_0$ ($c_0= \sqrt{5/4 \pi} \hbar / m \omega_0$). In $C_\chi$ \cite{grig3}, $\chi_0 c_0^2 = 96.3 A^{-5/3}$ MeV , which yields $\langle Q_{is,0} \rangle_g  \equiv \langle Q^p_0 \rangle_g + \langle Q^n_0\rangle_g = 0.18 \delta A^{5/3}$ fm$^2$, half the liquid drop estimate $A R_0^2 \delta/ \sqrt{5 \pi} = 0.36 \delta A^{5/3}$ fm$^2$, with $R_0= 1.2 A^{1/3}$ fm. }. 
Moreover, Eq. (2) is identical with the BWS condition for the two-rotor system,  giving (for $n=1$) the quantized angular oscillation amplitudes $a_\tau$ \cite{grig3}, 
$$a_\tau = \frac{1}{I_\tau} \sqrt{ \frac{2 \hbar I_r }{ \Omega_\chi} } ~~,~~\tau =p,n,~~I_r = \frac{I_p I_n}{I_p+I_n}~~, $$ 
while the excitation energy is 
${\cal E}_x \equiv \hbar \Omega_\chi = \hbar \sqrt{  C_\chi /  I_r}$. \\ \indent
To apply Eq. (3),  $\vert Z\rangle_\omega$  was approximated near $\vert g\rangle$
using a first order perturbative treatment of the cranking term. 
The "scissors-like" state (not normalized) obtained from Eq. (3) has the form 
\begin{equation}
\vert \Omega_\chi \rangle = \frac{1}{2 \hbar}(\frac{\Omega_\chi}{D} +1) \lbrack a_p  L^p_x -
a_n L^n_x \rbrack \vert g\rangle
\end{equation}
with $D= \vert \delta \vert \omega_0$. This state gives  the $B(M1)$ strength \cite{grig3} 
\begin{equation}
B(M1) = \frac{3}{4 \pi \hbar} (g_p-g_n)^2 I_r D ~\mu_N^2~~.
\end{equation}
The comparison with the experiment is complicated by the ambiguities of separating 
the orbital and the spin strengths, and defining a reasonable sum 
over fragments. Assuming the dominance of the orbital
strength at low energy, under 6 MeV in rare earths, and 4 MeV in
actinides, the data are well reproduced by Eq. (12) with the irrotational   
moments of inertia.   The range  of the energy ${\cal E}_x$  calculated by taking $I_r$ at the rigid or irrotational limits is not very large (below  2 MeV) and includes the 
data throughout all the mass regions investigated. 
\\ \indent
After normalization, the state $\vert \Omega_\chi\rangle$ becomes the same as the term
independent of spin and pairing from the state $\vert ROT\rangle$, constructed
before \cite{noj2} to represent microscopically the scissors modes.  
 However, this is not similar to an RPA state, because the "excitation operator"  $ a_p  L^p_x -
a_n L^n_x $ is Hermitian. A quasiboson operator may be obtained if 
the cranking wave function $\vert Z_\omega \rangle$ is related to $\vert g \rangle$ by a unitary transformation. This problem is not easy, but it was solved recently \cite{grig2} in terms of an unitary operator $U_\omega$ which relates the cranked anisotropic oscillator eigenstates to the eigenstates of a spherical harmonic oscillator. Thus, if  $\omega$ is not larger than $\omega_s \sqrt{3} /2$,  $\omega_{s}= \sqrt{( \omega^{2}_{y} + \omega^{2}_{z} )/2}$, then we have  \begin{equation}
  h_0 - \frac{ \delta}{3}  m \omega_0^2 (2 z^2 - x^2 -y^2) -
 \omega  l_x = U_\omega  h_s U^{-1}_\omega
\end{equation}
 with
\begin{equation}
U_\omega = e^{ -i \lambda c_{x}} \exp ( -i \sum_{k=x,y,z} \theta_{k} s_{2,k} )~~.
\end{equation}
In the left-hand side of Eq. (13) the first two terms correspond
to an anisotropic harmonic oscillator with the frequencies
$ \omega_{x}= \omega_y = \omega_0 \sqrt{1+ 2 \delta/3}$,
$\omega_z = \omega_0 \sqrt{1-4 \delta /3}$,  
while
$$ h_0= \sum_{k=x,y,z} \hbar \omega_0 (b_k^{\dagger} b_k +1/2), ~~~~ 
b^{\dagger}_{k} = \sqrt{ m \omega_0/ {2} \hbar} (x_{k} - i p_k /{m \omega_0})
$$
$$
h_s= \sum_{k=x,y,z} \hbar \Omega_k ( \tilde{b}_k^{\dagger} \tilde{b}_k +1/2),
 ~~~~ \tilde{b}^{\dagger}_{k}= \sqrt{ {m \omega_s }/{2} \hbar}(x_{k} -ip_k 
/{m \omega_s} )
$$
$$
l_x = i \hbar (b_y b^\dagger_z - b_z b^\dagger_y) ~~~~~
c_{x}= \tilde{b}^\dagger_{y} \tilde{b}_{z} + 
\tilde{b}^\dagger_{z} \tilde{b}_{y}, ~~~~~
s_{2,k} = i (( \tilde{b}^\dagger_k)^2 - ( \tilde{b}_k)^2)/4
$$ 
and  the parameters $\lambda$, $\theta_k$ of $U_\omega$ are
given by
$$
\tan 2 \lambda = 2 \omega / \omega_{s} \eta ~~~~~\sinh  \theta_{k} =
\omega_{s}(1-\omega_{k}^{2} / \omega_{s}^{2})/2 \Omega_{k} ~~~~~ 
 \eta=(\omega^{2}_{y} - \omega^{2}_{z})/2 \omega^{2}_{s} 
$$
$$
\Omega_{x}=
\omega_{x} ~~~~ \Omega_{y,z}^{2} =  (\omega_{s} + \epsilon_{y,z})^{2}
- (\omega_{s} \eta /2)^{2}
$$ 
with  $\epsilon_{y}= - \epsilon_{z} = \omega_{s} \eta / 2 \cos2 \lambda$.
 \\ \indent
The operators $s_{2,k}$, k=x,y,z, are the "squeezing" generators, 
and they produce the transition from a spherical to a deformed basis. This transition
corresponds to the "cranking" of $h_0$  by the term 
 $\delta m \omega_0^2 (2 z^2 - x^2 -y^2)/3$, proportional to 
$Q_0$. Therefore, the unitary operator $U_0$ with 
$\theta_k$ as variables may be used to generate trial manifolds 
for the treatment of the quadrupole vibrations\footnote{details can be found in Section III of the article  M. Grigorescu, N. C\^arjan, Dissipative shape dynamics in the sd shell, Phys. Rev. C {\bf 54} 706 (1996).} (${\cal E}_x \sim 2 \hbar \omega_0$). The commutation relations between $s_{2,k}$ and $b^\dagger_k b_k$ get closed to the su(1,1) ($\approx$sp(1,R)) algebra \cite{nieto}.  \\ \indent
The operator  $c_{x}$ generates the shift from a static frame to a frame rotating 
around the X-axis with the angular velocity $\omega$, and it appears as an 
"angle" operator conjugate to $l_x$.  Analog operators, $c_y$, $c_z$  are 
associated to $l_y$ and $l_z$, and by commutation
$c_x,c_y,c_z$ and $l_x,l_y,l_z$ generate an su(3) algebra\footnote{presented e.g. in 
 math-ph/0007033.}. Similarly,  $s_{2,x},s_{2,y},s_{2,z}$ and $l_x,l_y,l_z$ generate by commutation a gl(3,R) algebra. 
This rather complicated set of algebras is 
included in  the symplectic Lie algebra sp(3,R) \cite{rowe}.   \\ \indent
In the many-fermion case the one-body operators 
$h,h_0,h_s,l_k,c_k,s_k$ become  particle-hole operators, denoted 
$H,H_0,H_s,L_k,C_k,S_k$, and may be used to write 
 $\vert Z \rangle_\omega$ as  
$$\vert Z \rangle_\omega = e^{ -i \lambda C_{x}} \exp ( -i \sum_{k=x,y,z} \theta_{k} S_{2,k} )~ \vert g_s \rangle ~~.$$ 
This form is especially suited to study large amplitude vibrations, but corrections to Eq. (11) appear already in the linear approximation \cite{grig2}. If the dependence of $\theta_k$ on $\omega$ is neglected, then during the scissors vibration each $\vert Z\rangle_\omega$ in Eq. (9) changes in time only due to the factor of $C_x$, and Eq. (3) gives 
 $\vert \Omega_\chi\rangle = B^\dagger_\chi \vert g\rangle$, with \cite{grig2}
\begin{equation}
B^\dagger_\chi = \frac{1}{2 \hbar} \lbrack a^{p} { L}_{x}^{p}- a^{n} { L}_{x}^{n} -
\frac{ i \hbar \Omega_\chi }{\omega_{y} - \omega_{z}} (a^{p} { C}^{p}_{x} - 
a^{n} { C}^{n}_{x}) \rbrack~~.
\end{equation}
Unexpectedly, though  $a^{p,n}$ are the same angular amplitudes as in Eq. (11),  $B^\dagger_\chi$ is normalized in the RPA sense,  $\langle g \vert [B_\chi,B^\dagger_\chi] \vert g\rangle=1$.
In fact, if the particle-hole excitations between different oscillator shells are
neglected, then $ {\rm sgn}( \eta) i \hbar C_x \vert g\rangle \approx - L_x \vert g\rangle$,
and $B^\dagger_\chi \vert g\rangle$  reduces to Eq. (11).  \\ \indent
The operator $B^\dagger_\chi$ was obtained recently also by the canonical quantization of the TRM in relative coordinates \cite{noj3}, and it was proved to support the interpretation
of all low-lying orbital $1^+$ excitations as a scissors mode. \\ \indent
It is interesting to note that if the dependence of $\theta_k$ on $\omega$ is taken into 
account\footnote{M. Grigorescu, D. Rompf, W. Scheid, Dynamical effects of deformation in the coupled two-rotor system, Phys. Rev. C {\bf 57} 1218 (1998).},  then $B^\dagger_\chi$ will contain  operators from gl(3,R). Such terms are related to the excitation operator proposed long time ago by Hilton \cite{hilt}. Instead of constructing $B^\dagger_\chi$ within the "angular
momentum \& shift" algebra, su(3), he proposed a  combination within the "angular momentum \& squeezing" algebra, gl(3,R), between $l_y$ and the "shear" generator $ z p_x + x p_z 
\sim i (b^\dagger_z b^\dagger_x - b_z b_x)$. 
\subsection{The isovector Josephson oscillations } 
The combined effect of the proton-neutron interaction and breaking of the translational or rotational symmetries is related to the giant dipole resonance \cite{teller} or to the "scissors modes", respectively. However, there is one more important symmetry breaking in nuclei for which such type of isovector collective motion was not yet observed, and this appears when the nuclei are "superfluid".
\\ \indent
The ground state of a superfluid system accounts for the  pp correlations
produced by the pairing interaction, and is well approximated by a 
BCS function. For a single j-shell,
 the pairing Hamiltonian and the BCS function are
\begin{equation}
H_0 = \epsilon N - \frac{G}{4} P^\dagger P ~~,~~
\vert BCS\rangle(\varphi, \rho) = e^{(z P^\dagger - z^* P) } \vert 0\rangle~~,
\end{equation}
where $z = \rho e^{-i \varphi}$, $\varphi$ is the BCS "gauge" angle,
$ N= \sum_{m=-j}^j c^\dagger_{m} c_{m} $ 
is the particle-number  operator, and 
$$P^\dagger = \sum_{m=-j}^j (-1)^{j-m} c^\dagger_{m} c^\dagger_{-m}$$
is the pair creation operator.  \\ \indent
The angle $\varphi$ is not a 
constant, and a superfluid system in the ground state performs a
free gauge rotation with the angular velocity $\dot{ \varphi} = 2 \epsilon_F
/ \hbar$, twice the Fermi frequency. In a nucleus, the proton and neutron
systems are not isolated,  but change particles until the Fermi
energies $\epsilon_F^p$, $\epsilon_F^n$ become equal. Thus, we
may see this as an indication about the existence of a phenomenological
"gauge restoring interaction", which tends to fix the relative gauge
angle $\varphi_p - \varphi_n$ at a constant value. 
If this is true, and there is  an interaction  
between {\it pairs} of protons and neutrons, then
oscillations of the protons against neutrons in the BCS gauge space
should appear. \\ \indent
A Josephson-like proton-neutron interaction\footnote{considered also in M. Ger\c ceklio\u glu, A model for the doublets of the K$^\pi$=$0^+$ states in deformed nuclei, Acta Phys. Slovaca {\bf 52} 161 (2002).} 
$$H_{pn} = - \frac{\sigma}{4} (P^\dagger_p P_n + P_n^\dagger P_p) $$ 
may be related to the isospin symmetry breaking mean-field for
a  four particle interaction \cite{grig4}. The problem of the mean-field created by the 
four particle interaction is not new, but previously \cite{sol} the main 
interest was for terms  $ \sim P_p^\dagger P_n^\dagger$, assumed to 
represent $\alpha$ clusters, while terms like $ P_p^\dagger P_n$ were neglected.
Because $H_{pn}$ does not commute with the isospin $T_0 = (N_n -N_p)/2$,  
it produces an "isorotational" term $ k_\sigma (N-Z)^2$ in the total energy. 
This means a term in the symmetry energy $k_W (N-Z)^2$, ($k_W= 28/A$ MeV), from the Weisz\"acker mass formula  with a "dynamical" origin, beside the "kinematic" one determined by the Pauli
principle \cite{fesh}. Thus, $\sigma$ can be obtained by a fit of the symmetry energy produced by the Hamiltonian 
\begin{equation}
H = H_0^p + H_0^n + H_{pn} 
\end{equation}   
in a single j-shell. Considering the case of  $1d_{3/2}$ nuclei, 
$\sigma$ was  estimated to be $\approx 2.7 /A $ MeV \cite{grig4}. 
However, an approximation of  j-shells with high degeneracies
suggests a value about ten times larger \cite{grig5}.   \\ \indent
The interaction $H_{pn}$ contributes to the symmetry energy in 
all nuclei, but in the superfluid nuclei it produces also a restoring potential
$C_\sigma (\varphi_p - \varphi_n)^2/2$ for the BCS angles.  
This potential can be related to $H_{pn}$ by a treatment similar
to the one applied to the QQ interaction responsible for the orbital scissors
modes. Indeed, the BCS functions define symplectic manifolds ${\it
S}^{BCS}$ which can be  parameterized by $\varphi$ and $\rho$, or by
the canonical variables $\varphi$ and 
$$p = \langle BCS \vert N \vert BCS \rangle / 2 = (j + 1/2) \sin^2 2 \rho~~.$$ 
In these variables
 $$ \omega^{BCS}_{\varphi p} = 2 \hbar Im \langle  \partial_\varphi BCS( \varphi, 
\rho) \vert \partial_p BCS( \varphi, \rho)\rangle= \hbar \partial_{p}(\langle  BCS  
\vert N/2 \vert BCS\rangle) = \hbar~~. $$ For a proton-neutron system 
the trial manifold will be represented by the product
${\it S}^{BCS}_p \times {\it S}^{BCS}_n$, and the collective 
motion determined for $H$ by Eq. (1) shows the occurrence of isovector "gauge-angles" vibrations \cite{grig4}. For a half-filled shell, the fixed point in (1) corresponds to the ground state of $H$,
\begin{equation}
\vert g\rangle = e^{ \frac{\pi}{4} ( P^\dagger + N^\dagger - P - N)}
\vert 0\rangle
\end{equation} 
and the  gap parameter 
 $\Delta= G \langle  g \vert P^\dagger \vert g\rangle / 2$  is the same both for protons
and neutrons.  The  restoring potential has the constant 
$C_\sigma = 2 \sigma (\Delta / G)^2$, and the excitation energy for the isovector oscillations
is $E_\sigma = \hbar \Omega_\sigma = 4 \sqrt{2 k_W C_\sigma}$ \cite{grig4}. \\ \indent
The excitation operator defined by Eq. (3) has the form
\begin{equation}
B^\dagger_\sigma = \frac{1}{2j+1} \sqrt{\frac{E_\sigma}{\sigma}} \lbrack T_0  
 - \frac{ \sigma (2 j +1)}{4 E_\sigma} ( P^\dagger - P - N^\dagger
+ N) \rbrack
\end{equation}
and by accident, is the same as the one provided by the standard QRPA \cite{grig6}. However, 
the RPA vacuum defined by $B_\sigma \vert RPA\rangle=0$  exists only if $j + 1/2$ is an even integer \cite{grig5}.  \\ \indent 
The states generated by $B^\dagger$ are isovector monopoles, 
and correspond to  Josephson oscillations between the proton and
neutron superfluids\footnote{At $\sigma \ll G$ an odd-even effect in the total number of pairs can appear (M. Grigorescu, Low-lying excitations in superconducting bilayer systems, cond-mat/9904242, or High-Tc Update {\bf 13} No. 10, May 15 (1999), or Can. J. Phys. {\bf 78} 119 (2000)).}. Such oscillations might be excited by the Coulomb interaction in electron scattering \cite{grig5}, or by the current of pairs between the two superfluids produced in pion double 
charge-exchange (DCX) reactions. The importance of Josephson-type
correlations in  DCX reactions was proved first by
shell - model calculations \cite{auer}, suggesting  that the
"scissors modes" in gauge space discussed here are worth of experimental
investigation. 
\section{Summary} 
In this lecture I presented a microscopical approach to the collective motion,
based essentially on the time-dependent variational principle and
GIPQ requantization (Section 2.1),
but which is peculiar by the choice of the trial functions.
The trial manifolds are supposed to have  the phase-space (symplectic)
structure of a specific collective model, and for the symmetry breaking
nuclei are constructed using the cranking procedure \cite{grig2}.  
The Hamilton  equations of motion (Eq. (1)) appear  by
a constrained variational calculation in the Hilbert space,
rather than by semiclassical approximations ($\hbar \rightarrow 0$). 
This formalism includes the standard RPA or QRPA (Section 2.2), and was 
applied with success to the treatment of the collective isovector excitations
(Section 3). It solves the problems of the inertial parameters,
restoring force constants, and of the  microscopic analog
for a particular collective motion (here the scissors vibrations). Moreover, isovector
vibrations in the BCS gauge space of the superfluid nuclei were predicted. 
The variational formulation is appropriate to account for the
coupling between a quantum system and a thermal environment \cite{grig7}.
Therefore, the study of giant resonances using a Langevin form of
Eq. (1), with noise and memory-friction terms in the right-hand side, appears 
highly  interesting.

\end{document}